\documentclass[12pt]{article}
\usepackage{amsmath}
\usepackage{graphicx}
\usepackage{enumerate}
\usepackage{natbib}
\usepackage{url} 
\usepackage{amsfonts,amssymb}
\usepackage{float}
\usepackage{bm}
\usepackage{bbm}
\usepackage{algorithm}
\usepackage{algpseudocode}
\usepackage{amsthm}
\usepackage{lmodern}
\usepackage[colorlinks=true, allcolors=blue]{hyperref}
\usepackage{xcolor}
\usepackage{subfigure}
\newtheorem{assumption}{Assumption}
\newtheorem{lem}{Lemma}

\newtheorem{thm}{Theorem}

\newtheorem{cor}{Corollary}

\newtheorem{remark}{Remark}
\newcommand{\blind}{0}

\begin{document}

\def\spacingset#1{\renewcommand{\baselinestretch}%
{#1}\small\normalsize} \spacingset{1}


\if0\blind
{
  \title{\bf A General Stability Approach to False Discovery Rate Control}
  \author{Jiajun Sun$^1$, Zhanrui Cai$^2$ and Wei Zhong$^{3,1}$\thanks{Wei Zhong is the corresponding author. Email: wzhong@xmu.edu.cn.}\\
  $^{1}$ Department of Statistics \& Data Science, SOE, Xiamen University\\
  $^{2}$ Faculty of Business and Economics, University of Hong Kong\\
  $^{3}$ MOE Lab of Econometrics and WISE, Xiamen University}
  \date{\today}
  \maketitle
  
} \fi

\if1\blind
{
  \bigskip
  \bigskip
  \bigskip
  \begin{center}
    {\LARGE\bf A General Stability Approach to False Discovery Rate Control}
\end{center}
  \medskip
} \fi

\bigskip
\begin{abstract}
	
	Stability and reproducibility are essential considerations in various applications of statistical methods. False Discovery Rate (FDR) control methods are able to control false signals in scientific discoveries. However, many FDR control methods, such as Model-X knockoff and data-splitting approaches, yield unstable results due to the inherent randomness of the algorithms. To enhance the stability and reproducibility of statistical outcomes, we propose a general stability approach for FDR control in feature selection and multiple testing problems, named FDR Stabilizer. Taking feature selection as an example, our method first aggregates feature importance statistics obtained by multiple runs of the base FDR control procedure into a consensus ranking. Then, we construct a stabilized relaxed e-value for each feature and apply the e-BH procedure to these stabilized e-values to obtain the final selection set. We theoretically derive the finite-sample bounds for the FDR and the power of our method, and show that our method asymptotically controls the FDR without power loss. Moreover, we establish the stability of the proposed method, showing that the stabilized selection set converges to a deterministic limit as the number of repetitions increases. Extensive numerical experiments and applications to real datasets demonstrate that the proposed method generally outperforms existing alternatives.
	
\end{abstract}

\noindent%
\vfill

\newpage
\spacingset{1.8} 

\section{Introduction}

With the advancement of technology, statistical and machine learning communities have developed numerous methods for estimation and inference. While many of these methods aim to address complex scientific questions by optimizing specific objective functions or constructing test statistics, they may yield unstable results due to the inherent randomness in the algorithms or the data. For instance, in high-dimensional statistics, data splitting is commonly used to avoid double-dipping and to provide valid inference results \citep{Alessandro2019AOS, cai2024asymptotic}. However, random sample splitting can lead to divergent conclusions in practice when we only have access to finite sample sizes \citep{cai_model-free_2022}. \cite{yu2013stability} advocated that at the modeling stage, stability means acceptable consistency of model results relative to data or model perturbations, such as jackknife resampling, subsampling, bootstrap, etc. \cite{yu_veridical_2020} highlighted that stability is a core principle in veridical data science. 

In this paper, we aim to achieve algorithmic stability with False Discovery Rate (FDR) control in the multiple testing and variable selection framework. 
FDR control methods limit the proportion of false discoveries below a pre-specified level to ensure the validity of statistical results.
Let $R$ represent the total number of rejections and $V$ denote the number of false rejections. The FDR is defined as the expected proportion of false discoveries,
\[ FDR=\mathbb{E}[FDP],\quad FDP=\frac{V}{\max \{R,1\}}.\]
The concept of FDR was initially proposed in \cite{benjamini_controlling_1995}, where the authors proposed the famous Benjamini-Hochberg (BH) method by adjusting the rejection threshold for sorted $p$-values. Subsequently, this procedure was extended by \cite{benjamini2001control} to accommodate situations where all test statistics exhibit positive regression dependency. \cite{storey_strong_2004} introduced the q-value framework via estimation of the null proportion and proved FDR control under weak dependence. \cite{wu2008false} treated Markov dependence and provided validity conditions. \cite{clarke2009robustness} handled linear-process dependence. Recent research on FDR has focused on selecting important variables within general regression settings. Notably, \cite{barber_controlling_2015} proposed the knockoff method, which assesses variable importance by constructing exchangeable knockoff copies of the design matrix. This approach has since been further developed and generalized in various contexts, including Model-X knockoffs \citep{candes_panning_2018}, robust inference \citep{barber_robust_2020,fan2025ark}, multilayer knockoffs \citep{katsevich2019multilayer}, deep learning inference \citep{zhu2021deeplink}, and split knockoffs \citep{cao2024controlling,cao2024split}. z\cite{romano2020deep,bates2021metropolized} studied how to generate good knockoff copies. In another line of work, \cite{xing_controlling_2021, dai_false_2022,dai_scale_free_2023} used sample splitting and utilized the symmetry of mirror statistics, whose sampling distribution is asymptotically symmetric about zero for null features. The concept of using sample splitting to construct symmetric statistics was also explored in \cite{tong2023model}. Despite the successful applications of these approaches, some may yield unstable results due to the inherent randomness in the algorithms. To mitigate the instability caused by the inherent randomness in the algorithm, \cite{ren2023derandomized} proposed derandomized knockoffs via applying the e-BH procedure \citep{wang_false_2022} to the averaged e-values from multiple runs (see Section \ref{sec: backgrounds} for details). However, this procedure using averaged e-values is conservative in general. Moreover,
it lacks theoretical justification of 
power guarantee and stability enhancement.

\begin{figure}[ht]
\centering
\begin{minipage}{0.5\textwidth}
  \centering
  \includegraphics[width=\linewidth]{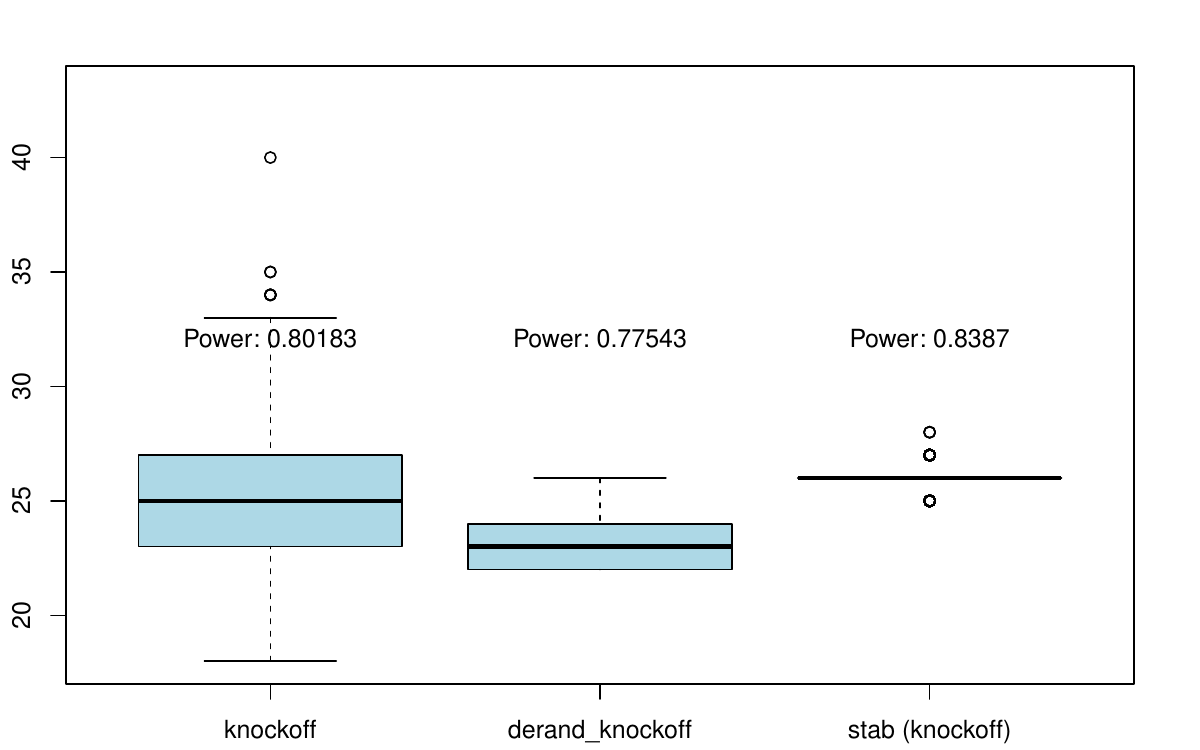}

\end{minipage}\hfill
\begin{minipage}{0.5\textwidth}
  \centering
  \includegraphics[width=\linewidth]{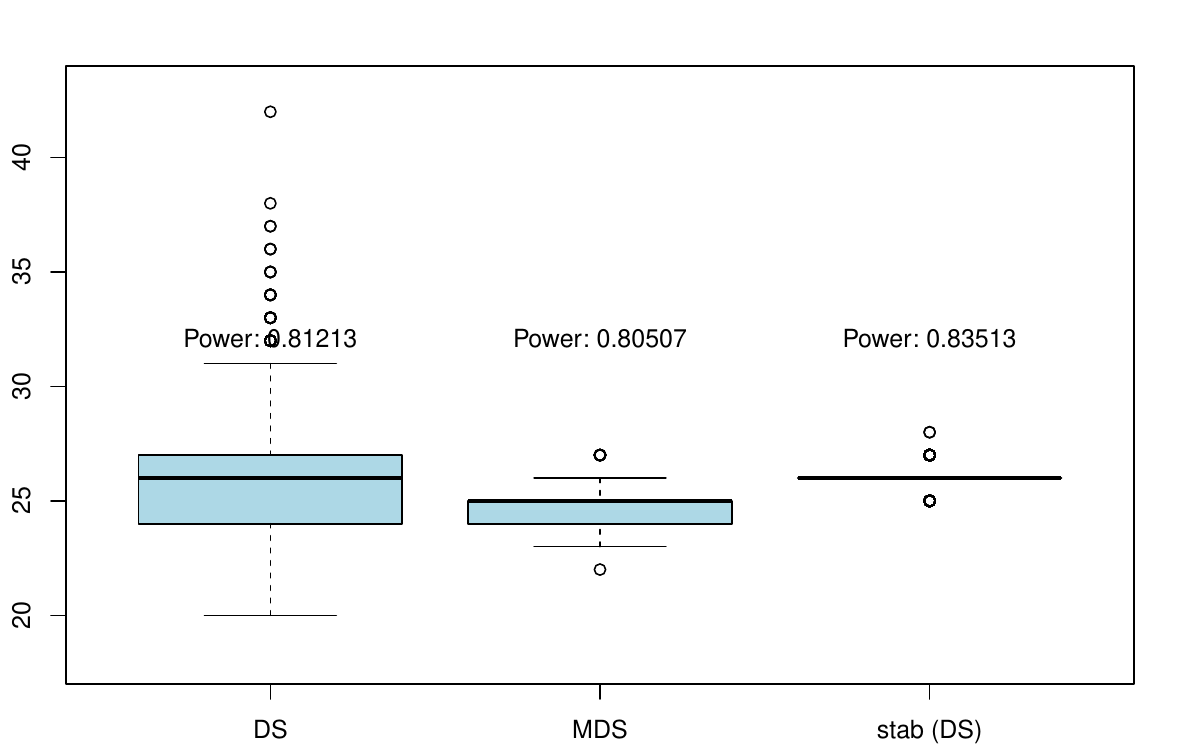}
\end{minipage}
\caption{Number of selected variables and average power for knockoff, derandomized knockoffs, and FDR Stabilizer (left); and  DS, MDS and FDR Stabilizer (right), based on 1000 repetitions. The target FDR level is $q = 0.1$.}
    \label{fig:motivation}
\end{figure}

In this paper, we propose a general stability framework for FDR control, named FDR Stabilizer, to stabilize the base FDR procedure with inherent randomness. Taking variable selection in high-dimensional data as an example, our method first aggregates the feature importance statistics returned by multiple runs of the base FDR procedure into a consensus ranking. Then, we construct a stabilized relaxed e-value for each feature and apply the e-BH procedure to these stabilized e-values to obtain the final set of selected variables. Next, to get the first insight into the motivation and excellent performance of the proposed method, we run a toy example to compare it with existing methods. Consider a linear model with $n=500$ samples and $p=200$ covariates, of which only $30$ covariates have non-zero coefficients drawn from Unif$(-1.5,-0.1)\cup$Unif$(0.1,1.5)$. The design matrix $X \sim N(0,\Sigma)$, where $\Sigma$ is a compound symmetry covariance matrix with $\Sigma_{ij}=0.5^{\mathbbm{1}(i\ne j)}$. We compare the number of selected variables using the methods in \cite{dai_false_2022} and \cite{barber_controlling_2015} (see the details in the simulation) with our stabilized versions in Figure \ref{fig:motivation} based on 1000 repetitions. On the left side, it is evident that the knockoff method \citep{barber_controlling_2015} and the derandomized knockoffs method \citep{ren2023derandomized} select a variable set whose size can vary substantially across different runs. In contrast, the proposed FDR Stabilizer method consistently identifies more significant variables than derandomized knockoffs with less fluctuation. On the right side, we observe that the Data Splitting (DS) method \citep{dai_false_2022} also yields unstable variable selection results. The Multiple Data Splitting (MDS) method can enhance the stability of DS but loss some power. However, our proposed method performs better, offering higher power and more stable results.

This paper makes several contributions. First, we propose a new general FDR Stabilizer framework which is applicable to a wide range of FDR procedures with inherent randomness, as long as multiple independent runs of the base FDR method are available. It is easy to implement and tuning-free. Second, we theoretically derive finite-sample bounds for the FDR and the power of our method, and show that our method asymptotically controls the FDR without power loss. Moreover, we prove the stability property of the proposed method that as the number of repetitions increases, the stabilized selection set converges to a deterministic set. Third, we conduct extensive simulation studies and real data analyses, demonstrating that our method consistently controls the FDR and achieves superior stability and power compared to existing alternatives. Finally, our framework can be naturally extended to FDR control methods without inherent randomness. In such cases, perturbation techniques such as subsampling can be incorporated to further enhance the stability.

This paper is organized as follows. Section \ref{sec: backgrounds} introduces the model settings and the background of the e-values. We propose the FDR Stabilizer methodology in Section \ref{sec: method}, followed by its theoretical properties in Section \ref{sec: theory}. Extensive numerical studies are presented in Section \ref{sec: simulation}. We also apply the proposed method to HIV Drug Resistance data and single-cell genomics data in Section \ref{sec: real data}. All the proofs and technical details are in the supplement.

\section{Model Settings and Backgrounds}\label{sec: backgrounds}

We aim to propose a general stability method for most existing multiple testing and variable selection problems with FDR control under various models. In multiple testing, the problem of interest is to test $p$ hypotheses simultaneously. Consider a collection of $p$ hypotheses $H_{01}, H_{02}, \ldots, H_{0p}$. Let $N\subset \{1,2,\dots,p\}$, $S=\{1,2,\dots,p\}\backslash N$ be the set of indices of true null hypotheses and true nonnull hypotheses, respectively. The cardinality of $S$, denoted by $s$, corresponds to the number of true nonnull hypotheses. Furthermore, let $\hat{S}$ denote the set of indices selected by a multiple testing procedure. In variable selection, we take a linear regression model $\mathbf{Y} = \mathbf{X} \boldsymbol{\beta} + \boldsymbol{\epsilon}$ as an illustrative example, where  $\mathbf{Y}= (Y_1, \ldots, Y_n)^{\top}$ is the response vector of sample size $n$,  $\mathbf{X}= (\mathbf{X}_{1}, \ldots, \mathbf{X}_{n})^{\top}$ is an $n\times p$ design matrix with covariates $\mathbf{X}_i=(X_{i1},\ldots, X_{ip})^{\top} \in \mathbb{R}^p$,  $\boldsymbol{\epsilon}=(\epsilon_1,\ldots, \epsilon_n)^{\top}$ is a vector of random noises, and $\boldsymbol{\beta} = (\beta_1, \ldots,\beta_p)^{\top}$ is the vector of regression coefficients. In this case, $S=\{k:\beta_k \ne 0\}$ and $N=\{k:\beta_k=0\}$. The objective is to estimate the relevant variable set $S$ with FDR control in high dimensions where $p$ is generally greater than the sample size $n$.

\subsection{E-values and the e-BH procedure}
E-values \citep{shafer2011test,vovk_e-values_2021,wang_false_2022,ramdas2024hypothesis,koning2024continuous} have emerged as a powerful tool for multiple testing, allowing FDR control under arbitrary dependence. To be concrete, a non-negative random variable $E$ is called an e-value if $\mathbb{E}[E] \leq 1$ under the null hypothesis. E-values are summaries of evidence against a null hypothesis $H_0$, where we reject $H_0$ for large e-values. Formally, if we choose a significance $\alpha\in (0,1)$ and reject $H_0$ whenever $e\geq 1/\alpha$, Markov’s inequality ensures that
\[ \mathbb{P}_{H_0}(e\geq 1/\alpha)\leq \alpha\mathbb{E}_{H_0}(e)\leq \alpha. \]

In the context of multiple tests, each null hypothesis $H_{0i}$ is associated with an e-value $e_i$.  \cite{wang_false_2022} proposed an e-BH procedure for FDR control based on e-values.
The e-BH procedure operates similarly to the widely used BH procedure and determines the rejection set based on the following equation,
\[ \hat{S}_{\mathrm{ebh}}=\Big\{i:e_i\geq\frac{p}{q\hat{k}}\Big\}, \mathrm{~where~}\hat{k}=\max\Big\{k\in[p]:e_{(k)}\geq\frac{p}{q k}\Big\}, \]
where $e_{(1)}\geq\cdots\geq e_{(p)}$ are the order statistics of $e_i$'s. \cite{wang_false_2022} demonstrated that the e-BH procedure controls the FDR at the desired level for any dependence structure among the e-values. To understand the e-BH procedure, we now briefly go through the proof of its FDR control. We can write the FDR of e-BH as
\begin{equation}
\label{eq: fdr proof}
    FDR=\sum_{i\in N}\mathbb{E}\left(\frac{\mathbbm{1}\{e_i\geq \frac{p}{q|\hat{S}_{\mathrm{ebh}}|} \}}{|\hat{S}_{\mathrm{ebh}}|\vee 1}\right)\overset{(a)}{\leq}\sum_{i\in N}\mathbb{E}\left(\frac{e_i \frac{q|\hat{S}_{\mathrm{ebh}}|}{p}}{|\hat{S}_{\mathrm{ebh}}|\vee 1}\right)\leq\frac{q}{p}\sum_{i\in N}\mathbb{E}\left(e_i\right)\overset{(b)}{\leq}q,
\end{equation}
where step ($a$) follows from the deterministic inequality $\mathbbm{1}\{X\geq t\} \leq X/t$ for any $t > 0$, and step ($b$) is due to the definition of e-values. It is sufficient for FDR control to require that $\sum_{i\in N}\mathbb{E}\left(e_i\right)\leq p$. This observation naturally motivates the definition of relaxed e-values (also referred to as compound e-values in \cite{ignatiadis2024values}), defined as non-negative random variables $e_1,\dots,e_p$ satisfying $\sum_{i\in N}\mathbb{E}(e_i)\le p$. Applying the e-BH procedure to relaxed e-values controls FDR at level $q$.

\subsection{Derandomized knockoffs}

To reduce the randomness of the Model-X knockoffs filter \citep{candes_panning_2018}, \cite{ren2023derandomized} proposed derandomized knockoffs. They first established an explicit connection between the Model-X knockoffs and the e-BH procedure by defining relaxed e-values
\begin{equation}
    \label{eq: rb}
    e_i^{\mathrm{RB}}=p\cdot\frac{\mathbbm{1}\{W_i\geq \tau\}}{1+\sum_{k\in [p]}\mathbbm{1}\{W_k\leq -\tau\}},
\end{equation}
where $W_i$ is the feature importance statistics for $X_i$ and $\tau=\inf\left\{t>0:\frac{1 + \sum_{k=1}^p \mathbbm{1}\{W_k \leq -t\}}{\sum_{k=1}^p \mathbbm{1}\{W_k \geq t\}}\leq q\right\}$. By the FDR control property of Model-X knockoffs, it can be shown that $e_1,\dots, e_p$ are relaxed e-values. Building on this connection, \cite{ren2023derandomized} repeatedly run the knockoff algorithm to obtain multiple sets of relaxed e-values, and then apply the e-BH procedure to the averaged e-values to produce a stable final selection. Specifically, fix a tuning parameter $q_{\mathrm{kn}} \in (0,1)$ and generate $M$ knockoff copies 
$\widetilde{\mathbf{X}}^{(1)}, \dots, \widetilde{\mathbf{X}}^{(M)}$. 
For the $m$th knockoff replicate, compute feature importance statistics 
$W^{(m)} = \mathcal{W}([\mathbf{X}, \widetilde{\mathbf{X}}^{(m)}], \mathbf{Y})$ 
and determine a threshold $\tau^{(m)}$ by 
\(\tau^{(m)} = \inf \left\{ t > 0 : 
\frac{1 + \sum_{k \in [p]} \mathbbm{1}\{W^{(m)}_k \le -t\}}
     {\sum_{k \in [p]} \mathbbm{1}\{W^{(m)}_k \ge t\}}
\le q_{\mathrm{kn}} \right\}.\) 
Based on this threshold, the relaxed e-values for the $m$th knockoff run are defined as
\[e^{(m)}_i = p \cdot 
\frac{\mathbbm{1}\{W^{(m)}_i \ge \tau^{(m)}\}}
{1 + \sum_{k \in [p]} \mathbbm{1}\{W^{(m)}_k \le -\tau^{(m)}\}},\quad i=1,\dots,p.\]

For each feature $i$, a derandomized e-value can be obtained by averaging the relaxed e-values $e^{(m)}_i$ computed in each of the $M$ runs of the knockoff filter: $e^{\mathrm{avg}}_i:=\frac{1}{M}\sum_{m=1}^M e^{(m)}_i$.
This averaged e-value reduces the variability inherent to individual knockoff runs. Because averaging preserves the validity of relaxed e-values, applying the e-BH procedure to $(e^{\mathrm{avg}}_1,\dots,e^{\mathrm{avg}}_p)$ yields a rejection set that controls the FDR at the target level $q$. In the recent work of \cite{li2025note}, who established a general framework showing that a broad class of FDR-controlling procedures, including BH, Storey’s procedure, and the knockoff filter, can each be expressed as an e-BH procedure with appropriately defined relaxed e-values. The construction in \cite{ren2023derandomized} can be viewed as one concrete instantiation of this general principle.

However, at least two limitations of derandomized knockoffs have not been resolved well. First, how to choose the optimal value of parameter $q_{\mathrm{kn}}$ is unknown, which directly affects the power. Second, the power of the derandomized knockoff procedure can be severely influenced by correlations among covariates. One key reason is that step ($a$) in equation \eqref{eq: fdr proof} is a major source of discrepancy between the realized FDR of e-BH and its nominal level $q$. Notably, step ($a$) is tight if and only if each $e_i \in \{0, \frac{p}{q|\hat{S}_{\mathrm{ebh}}|}\}$. For the relaxed e-values $e_i^{\mathrm{RB}}$ derived from a single knockoff run, this step is tight, whereas averaging across replicates to obtain $e_i^{\mathrm{avg}}$ breaks this tightness, leading to more conservative FDR control and, consequently, a loss of power.


\section{Methodology: FDR Stabilizer}\label{sec: method}

\subsection{Stabilized e-value}

To broaden the applicability of our methodology, we propose a general and flexible definition of relaxed e-values that is not tied to a specific procedure. Without loss of generality, let $\bm{T} = (T_1, \dots, T_p)$ denote feature importance statistics, 
where a larger value of $T_i$ indicates greater evidence against the null hypothesis for the $i$th feature. For instance, in variable selection, one may take $T_i = W_i$, the knockoff statistics. We denote $\pi(i)$ as the rank of $T_i$ in the list $(T_1,\dots,T_p)$. A lower rank of $\pi(i)$ indicates more preference for the $i$th feature.
A general relaxed e-value for any FDR control method with level $q$ is defined as 
\begin{equation}
    \label{eq_e20}
    e_i = p\frac{\mathbbm{1}(\pi(i)\leq \hat{s})}{q(\hat{s}\vee 1)},
\end{equation}
where $\hat{s}$ is the number of selected variables, determined implicitly by the underlying FDR-controlling base procedure operating at the target level $q$. If the base procedure controls the FDR, i.e., $\mathbb{E}(\sum_{i\in N}\frac{\mathbbm{1}(\pi(i)\leq \hat{s})}{\hat{s}\vee 1})\leq q$, by times $p/q$ on both sides, it immediately implies that the defined e-value in equation \eqref{eq_e20} is a relaxed e-value, i.e. $\mathbb{E}\left(\sum_{i\in N}e_i\right)\leq p$. In the denominator of the general relaxed e-value, the number of false discoveries is approximated by $q(\hat{s}\vee 1)$ so that the definition does not rely on any specific structural properties of the underlying FDR-controlling procedure. In contrast, the relaxed e-values defined by equation \eqref{eq: rb} use the intrinsic symmetry properties of knockoff statistics to estimate the number of false discoveries. This estimate may not apply in general cases, such as when only the p-value is available.

As we mentioned before, many FDR control methods rely on randomized algorithms, which can yield unstable selection outcomes with a single run, such as Model-X knockoff \citep{candes_panning_2018}, Data Splitting \citep{dai_false_2022,dai_scale_free_2023,du_false_2023}, Gaussian Mirror \citep{xing_controlling_2021}, and among others.
To reduce the randomness of the FDR control procedures,  
we run the base variable selection procedure $M$ times with a specified FDR level $q$. For each $m=1,\ldots, M$, 
we let $\bm{T}^{(m)}=(T_1^{(m)},\dots,T_p^{(m)})$,
$\bm{\pi}_m=(\pi_m(1),\pi_m(2),\cdots,\pi_m(p))$,
$\hat{s}_m$ denote the feature importance statistics, the rank vector, and the number of selected variables at the $m$th run, respectively. Then, the general relaxed e-value at the $m$th run is
\begin{equation}
    \label{eq_e2}
    e_i^{(m)} = p\frac{\mathbbm{1}(\pi_m(i)\leq \hat{s}_m)}{q(\hat{s}_m\vee 1)}.
\end{equation}
How to aggregate the relaxed e-values over $M$ runs is crucial to guarantee both the power of variable selection and FDR control as well as the stability of the final result. Averaging relaxed e-values across multiple runs, as done in \cite{ren2023derandomized}, loosens step ($a$) in the FDR proof of e-BH \eqref{eq: fdr proof}, leading to a conservative procedure and a noticeable loss of power, particularly under strong covariate correlations. 
To retain the tightness of step ($a$) and thus preserve the power of the selection procedure, we propose a componentwise stabilization strategy that directly aggregates feature-level importance scores before constructing e-values. Specifically, we define the stabilized relaxed e-value as
\begin{equation}
    \label{eq_e_stab}
    e_i^{\mathrm{stab}}=p\frac{\mathbbm{1}(\tilde{\pi}(i)\leq \bar{s})}{q(\bar{s} \vee 1)}, \quad i=1,2,\dots,p,
\end{equation}
where $\tilde{\pi}(i)$ is the rank of an aggregation statistic 
$g({T}_i^{(1)},{T}_i^{(2)},\cdots,{T}_i^{(M)})$ for the $i$th variable and $\bar{s}=\lceil\frac{1}{M}\sum_{m=1}^M \hat{s}_m\rceil$ is the average number of the selected variables over $M$ runs, where $\lceil \cdot \rceil$ denotes the ceiling operator. Here, $g$ is a user-chosen function, e.g. mean or median, that combines the $M$ statistics for variable $i$ into one scalar. We assume no ties in the aggregation statistics for simplicity.
Comparing $e_i^{\mathrm{stab}}$ with $e_i$ in equation \eqref{eq_e20}, the rank ${\pi}(i)$ of the $i$th feature importance statistic is replaced by its stable aggregation version $\tilde{\pi}(i)$ over $M$ runs and the number of selected variables $\hat{s}$ is replaced by the average number of the selected variables over $M$ runs. Our design aims to tighten step ($a$) in the FDR proof of e-BH \eqref{eq: fdr proof}, thereby reducing the conservativeness observed in \cite{ren2023derandomized} and improving the power. This componentwise stabilization is not only able to make the whole FDR procedure stable but also reduce the influence of strong correlations among variables and avoid power loss. In addition, we also note that if the base method only outputs the final selection sets rather than explicit statistics, our framework still applies: in this case, one can regard all selected hypotheses as tied at the top of the ranking. Aggregation can then be based on measures such as selection probabilities across runs, which are consistent with our general framework.

It is worth noting that the conditional calibration framework of \cite{lee2024boosting} is a highly relevant recent contribution that aims to improve the power of average e-BH while retaining finite-sample FDR control. By conditioning on suitable sufficient statistics and using resampling to calibrate null distributions, conditional calibration can effectively mitigate the ``threshold phenomenon'' inherent in derandomized knockoff, leading to more powerful testing decisions in certain settings, such as sparse signals setting. However, conditional calibration requires access to sufficient statistics and null-preserving sampling mechanisms, which may not be straightforward to derive for widely used procedures such as data splitting; in contrast, our method only assumes the ability to repeatedly run the base procedure and then perform a simple rank-based aggregation, which makes it directly applicable in a much broader range of settings.

\subsection{FDR Stabilizer and examples}

Once we obtain the stabilized e-values over $M$ runs,
we apply the e-BH procedure at level $q$ to the stabilized e-values $(e_1^{\mathrm{stab}}, e_2^{\mathrm{stab}}, \ldots, e_p^{\mathrm{stab}})$ to obtain the final stabilized set of selected variables, denoted by $\hat{S}^{\mathrm{stab}}_{q}$. This procedure, which we call the FDR Stabilizer, is summarized in Algorithm \ref{algo1}.

\begin{algorithm}[ht!]
  \caption{FDR Stabilizer}
  \label{algo1}
  \begin{algorithmic}
    \Require
        data $(\mathbf{X},\mathbf{Y})$, the FDR control level $q$, the base variable selection procedure $\hat{S}_q$, the number of independent realizations $M$;
    
    \Ensure
        the final stabilized selection set $\hat{S}^{\mathrm{stab}}_{q}$;

    \begin{enumerate}
        \item Run the base variable selection procedure $M$ times independently on the dataset $(\mathbf{X},\mathbf{Y})$ to obtain the feature importance statistics        
        $\bm{T}^{(m)}=(T_1^{(m)},\dots,T_p^{(m)})$ 
and the number of selected variable $\hat{s}_m$,  
$m=1,\dots,M$;
        \item Compute $\tilde{\pi}(i)$, the rank of an aggregation statistic 
$g({T}_i^{(1)},{T}_i^{(2)},\cdots,{T}_i^{(M)})$, for $i=1,2,\ldots, p$ and $\bar{s}=\lceil\frac{1}{M}\sum_{m=1}^M \hat{s}_m\rceil$, the average number of the selected variables over $M$ runs. Then, compute the stabilized e-values $e_i^{\mathrm{stab}}, i=1,\dots,p$ according to equation \eqref{eq_e_stab};
        \item Compute $\widehat{k}=\max\left\{k:e_{(k)}^\mathrm{stab}\geq p/qk\right\}$. If this set is empty, we take $\widehat{k}=0$;
        \item Compute the final stabilized selection set $\hat{S}^{\mathrm{stab}}_{q} = \{i:e_{i}^\mathrm{stab}\geq p/q\widehat{k}\}$.
    \end{enumerate}
  \end{algorithmic}
\end{algorithm}

We first establish that e-BH applying on
$\bm{e}^{\mathrm{stab}}$ is algebraically equivalent to selecting the top-$\bar{s}$ features ranked by the aggregated statistics in the following Lemma \ref{thm_stl}. 
\begin{lem}
    \label{thm_stl}
    Let $\hat{S}^{\mathrm{stab}}_q$ be the final set of the selected variables via applying the e-BH procedure to stabilized e-values $e_{1}^{\mathrm{stab}},\ldots,e_{p}^{\mathrm{stab}}$ at the desired FDR control level $q$. Then, 
    $\hat{S}^{\mathrm{stab}}_q=\{i: \tilde{\pi}(i)\leq \bar{s}\}.$
\end{lem}

The choice of the aggregation function $g({T}_i^{(1)},{T}_i^{(2)},\cdots,{T}_i^{(M)})$ is a key component of our framework and is flexible. A variety of simple yet effective options are available. For example, the mean or the median of $({T}_i^{(1)},{T}_i^{(2)},\cdots,{T}_i^{(M)})$, the selection probability $\Pi_i=\frac1M\sum_{m=1}^M\mathbbm{1}\{\pi_m(i)\leq \hat{s}_m\}$, the average of e-values $e_i^{\mathrm{avg}} = \frac{1}{M}\sum_{m=1}^M e_i^{(m)}$. The flexibility in the choice of $g$ allows our framework to be adapted to a range of application scenarios. For example, simple averages or medians are computationally efficient in moderate dimensions, while probabilistic rank aggregation methods can further refine rankings in challenging regimes with strong correlations or weak signals, thereby improving both stability and power. 

To systematically evaluate the performance of different aggregation functions $g$, we perform a simulation comparing seven aggregation strategies: the mean of $\{T_i^{(m)}\}_{m=1}^M$ (mean), the median of $\{T_i^{(m)}\}_{m=1}^M$ (median), the mean of $\{\pi_m\}_{m=1}^M$ (rank\_mean), the selection probability $\Pi_i = \frac{1}{M} \sum_{m=1}^M \mathbbm{1}\{\pi_m(i)\leq \hat{s}_m\}$ (sel\_prob), the average of e-values $e_i^{\mathrm{avg}} = \frac{1}{M}\sum_{m=1}^M e_i^{(m)}$ (e\_avg), the consensus ranking obtained by fitting a Mallows model computed from $\{\pi_m\}_{m=1}^M$ (MM), the consensus ranking obtained by fitting an Extended Mallows model \citep{RankAggregation} computed from $\{\pi_m\}_{m=1}^M$ (EMM). 

\begin{figure}[ht!]
        \centering
        \includegraphics[width=1\linewidth]{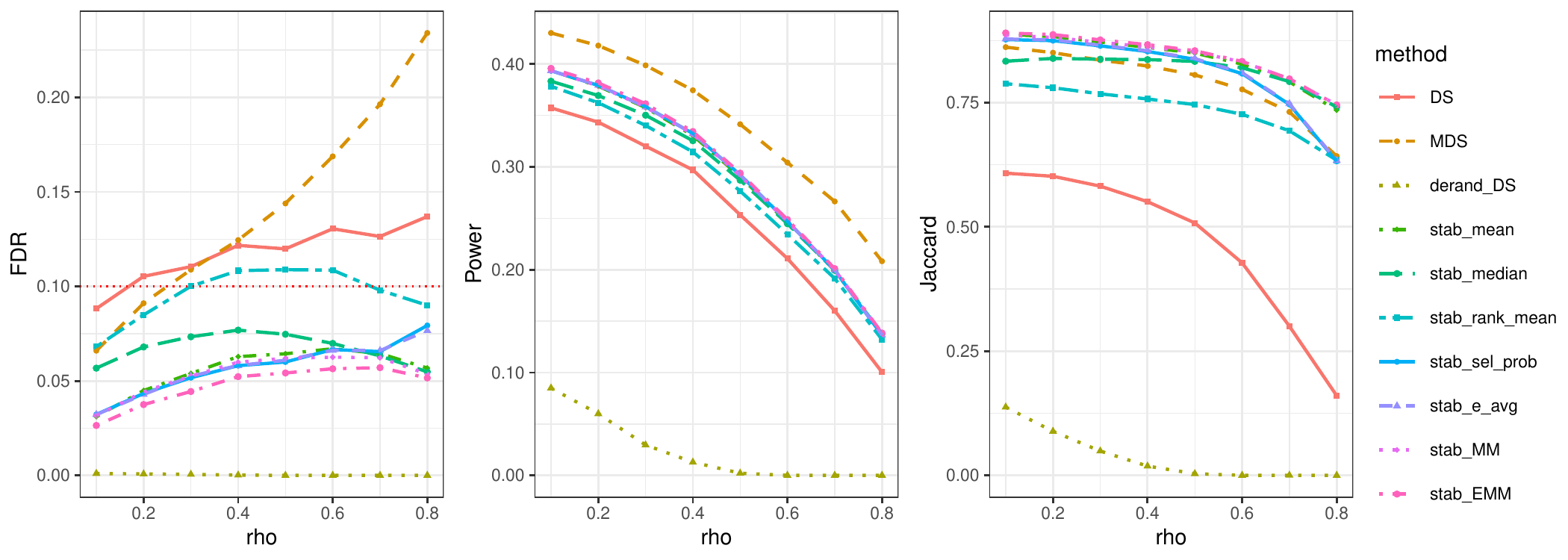}
        \includegraphics[width=1\linewidth]{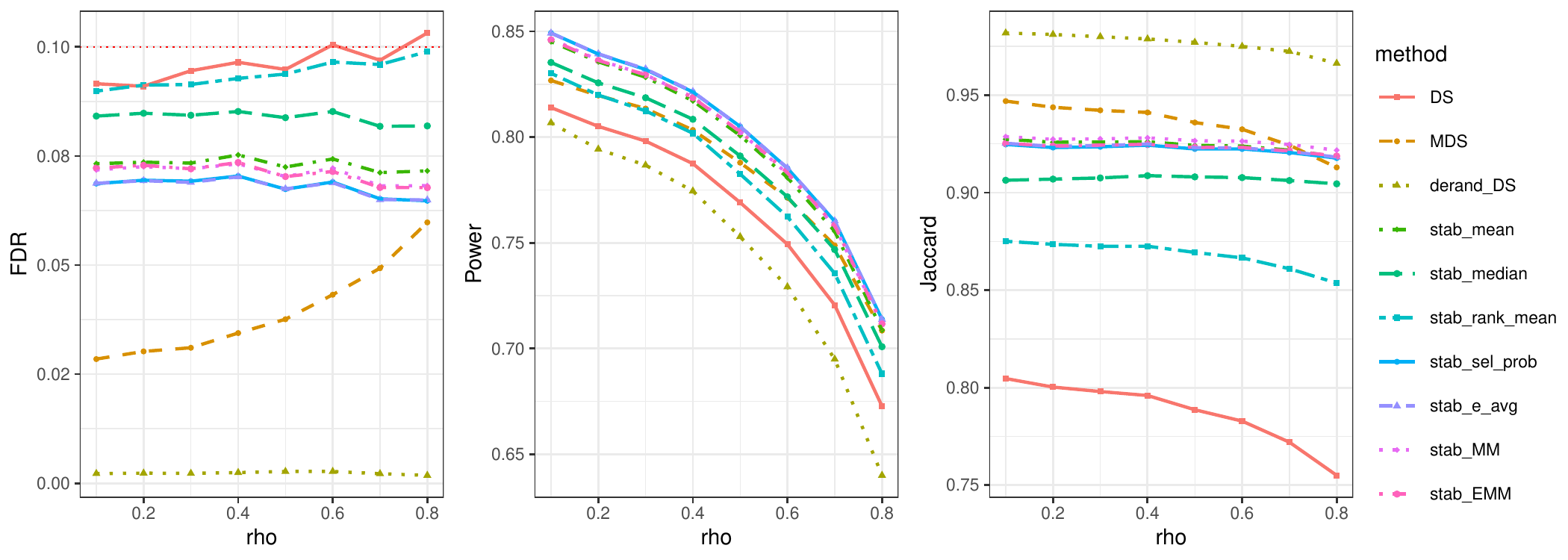}
        
        \caption{Empirical FDR, power and jaccard index performances of different methods when $n=800$, $p=2000$. With $\delta=2$ and $\delta=8$ for the top and bottom panels, respectively. $\Sigma$ is the blockwise diagonal Toeplitz covariance matrix where $\rho$ varies from 0.1 to 0.8. The specified FDR control level is $q = 0.1$.}
        \label{fig: g}
    \end{figure}

To assess their impact on stability and power, we adopt the simulation settings introduced in Section \ref{sec: simulation}, using exactly the same data-generating models and parameter configurations described there, ensuring comparability with the main simulation study. We use the Jaccard index to assess the stability of the method, whose definition is provided in Section S3.3 of the Supplementary Material. We take the DS procedure as the base FDR-controlling method and compare our FDR Stabilizer framework with two representative stabilization approaches: MDS, and a derandomized knockoff variant adapted to the DS setting, referred to as derand\_DS. Figure \ref{fig: g} presents the simulation results comparing different aggregation strategies, where our method is denoted as stab followed by a suffix indicating the aggregation strategy used for $g$, e.g., stab\_mean, stab\_median, or stab\_EMM. The results in Figure \ref{fig: g} empirically demonstrate clear differences among stabilization strategies under varying signal strengths and correlation settings. In weak-signal scenarios, the derand\_DS method suffers from markedly low power, while MDS frequently exceeds the target FDR level; both approaches also exhibit poor stability, reflected in low Jaccard indices. By contrast, our proposed stability procedures with different choices of aggregation function $g$ consistently achieves nominal FDR control, substantially higher power, and improved stability. Under strong signals, all methods attain satisfactory FDR control and high power, but Stab achieves the highest power overall; all Jaccard index is slightly lower than that of MDS and derand\_DS, though all are close to one. In practice, the results obtained with different choices of $g$ are very close. For ease of comparison with competing methods, this paper mainly uses the rank of the averaged e-values as a special case of $\tilde{\pi}$. 

Next, we present several specific examples to illustrate the applicability of the proposed FDR Stabilizer procedure to make the existing FDR control methods more stable and powerful. Numerical comparisons between the existing methods and their stabilized versions based on our FDR Stabilizer procedure are presented in Section \ref{sec: simulation}.

\vspace{0.4cm}
\noindent\textbf{Example 1: FDR Stabilizer + MBH.}
\cite{meinshausen_p_2009} proposed a multisplit method for assigning statistical significance and constructing conservative $p$-values for variable selection under high-dimensional data based on the BH procedure, denoted as MBH.
To be specific, the data are randomly split into two parts with the same size. The first part is used to estimate the set of active predictors and then ordinary least squares is employed in the second part to obtain the $p$-values. After $M$ data splits, each variable has a total of $M$ $p$-values, $\bm{P}_i=(P_i^{(1)},\dots,P_i^{(M)})$, for $i=1,2,\ldots, p$. \cite{meinshausen_p_2009} proposed an adjusted $p$-value for each variable based on the quantile of $\bm{P}_i$ across multiple splits to deal with the ``$p$-value lottery" problem of the single split. Then, the BH procedure is applied to these aggregated $p$-values for the final variable selection with FDR control.
Our FDR Stabilizer method offers an alternative aggregation scheme that could yield higher power. Specifically, we perform $M$ independent data splits and set the feature importance statistics $\bm{T}^{(m)}=-\bm{P}^{(m)}=(-P_1^{(m)},-P_2^{(m)},\dots,-P_p^{(m)})$ for the $m$-th run.  Then, we calculate stabilized e-values $e_{i,\mathrm{bh}}^{\mathrm{stab}}, i=1,\dots,p$ according to equation \eqref{eq_e_stab} and run the e-BH procedure on $\bm{e}_{\mathrm{bh}}^{\mathrm{stab}}$ to obtain the stabilized selection set $\hat{S}_{q,\mathrm{bh}}^{\mathrm{stab}}$. 

\vspace{0.4cm}
\noindent\textbf{Example 2: FDR Stabilizer + knockoffs.} 
\cite{ren2023derandomized} defined a relaxed e-value by
\begin{equation}
    \label{eq_ren}
    e_i^{\mathrm{RB}}:=p\frac{\mathbbm{1}\{T_i\geq \tau_q\}}{1+\sum_{i=1}^p \mathbbm{1}\{T_i\leq -\tau_q\}}, \quad  i=1,2,\dots,p,
\end{equation}
where $\tau_q = \inf\left\{t>0:\frac{1+\sum_{i=1}^p\mathbbm1\{T_i\leq-t\}}{\sum_{i=1}^p\mathbbm1\{T_i\geq t\}}\leq q\right\}$.
\cite{ren2023derandomized} proposed the derandomized knockoffs to  reduce the randomness inherent to Model-X knockoffs by applying the e-BH procedure \citep{wang_false_2022} to the averaged e-values over multiple runs. Different from the aggregation method in  \cite{ren2023derandomized}, we adopt the idea of componentwise stabilization to obtain new stabilized e-values
according to equation \eqref{eq_e_stab}, where the aggregation function $g({T}_i^{(1)},{T}_i^{(2)},\cdots,{T}_i^{(M)})$ is set to be 
the average of e-values $e_i^{\mathrm{avg}} = \frac{1}{M}\sum_{m=1}^M e_i^{(m)}$ as a special case. 
Once we obtain stabilized e-values $e_{i,\mathrm{kn}}^{\mathrm{stab}}, i=1,\dots,p$, we run the e-BH procedure to obtain the stabilized selection set $\hat{S}_{q,\mathrm{kn}}^{\mathrm{stab}}$. We note that recent extensions of knockoffs, such as Split Knockoffs for structured sparsity and directional FDR control \citep{cao2024controlling,cao2024split}, can also be used as base procedures within our stability framework.

 \vspace{0.4cm}
\noindent\textbf{Example 3: FDR Stabilizer + data splitting.}  
The data-splitting technique has been proposed by \cite{dai_false_2022,dai_scale_free_2023} and \cite{du_false_2023} to select relevant variables with FDR control. To be specific, the data are first randomly split into two parts with equal size and 
two independent sets of regression coefficient estimates $\hat{\bm{\beta}}^{(1)}$ and $\hat{\bm{\beta}}^{(1)}$ are obtained  potentially with two different fitting algorithms for two parts of data. Then, the feature importance statistics, also called mirror statistics, $\bm{T}=(T_1,T_2,\dots,T_p)$ with the form of $T_i= \mathrm{sign}(\hat{\beta}_{i}^{(1)}\hat{\beta}_{i}^{(2)})f(|\hat{\beta}_{i}^{(1)}|,|\hat{\beta}_{i}^{(2)}|)$ are constructed, where $f(u,v)$ is a nonnegative, exchangeable, and monotonically increasing function defined for nonnegative $u$ and $v$, for example, $f(u,v)=uv$. 
The sampling distribution of  mirror statistics is symmetric about zero for any null variable and has a positive mean for a relevant variable. Subsequently, variable selection with FDR control is performed similar to the knockoff method. To reduce the randomness of data splits, \cite{dai_false_2022,dai_scale_free_2023} suggested using the estimated inclusion rate based on MDS to select relevant variables, that is, $\hat{I}_i=\frac1M\sum_{m=1}^M\frac{\mathbbm{1}(i\in\hat{S}^{(m)})}{\hat{s}_m\vee1}.$
Intuitively, the larger the inclusion rate is, the more likely the corresponding variable is relevant. For ease of comparison, we use the rank of the mean of the e-values as a special case of $\tilde{\pi}$, which is in fact a constant multiple of the inclusion rate. 
Different from their aggregation method, we run the e-BH procedure on $\bm{e}_{\mathrm{ds}}^{\mathrm{stab}}$ to obtain the stabilized selection set $\hat{S}_{q,\mathrm{ds}}^{\mathrm{stab}}$. 
Our FDR Stabilizer method offers an alternative aggregation scheme to make the final selection set more stable and powerful, especially when the variables are strongly correlated, which has been demonstrated via numerical comparisons in Section \ref{sec: simulation}.

\section{Theoretical Properties}\label{sec: theory}
We study the theoretical properties of the proposed FDR Stabilizer method, including FDR control and power guarantee, and stability enhancement. The construction of $\bm{e}^{\mathrm{stab}}$ comes at a cost: it may no longer satisfy the relaxed e-value property, so step ($b$) of the proof is not automatically guaranteed. To ensure that step ($b$) holds, we need some mild assumptions. Before the theoretical results, we first introduce some notation. For each variable $i\in\{1,\dots,p\}$ and each run $m=1,\dots,M$, we define \(\mu_i:=\mathbb{E}(T_i^{(m)}\mid \mathbf X,\mathbf Y)\), 
\(g_i:=g\big(T_i^{(1)},\dots,T_i^{(M)}\big)\), and \(\eta_i:=\mathbb{E}(g_i\mid \mathbf X,\mathbf Y)\).

\begin{assumption}
\label{ass: 1}
    There exist sequences $\delta_{np},b_{np}\geq 0$ such that
  \[
  \frac{1}{|N||S|}\sum_{i\in N, j\in S}\mathbb{P}(\mu_i> \mu_j) \leq \delta_{np},\quad \frac{1}{|N||S|}\sum_{i\in N, j\in S}\mathbb{P}(\eta_j<\eta_i, \mu_j>\mu_i)\leq b_{np}.
  \]
\end{assumption}

\begin{assumption}
\label{ass: 2}
    There exist constants $a>0$ and $K_g>0$ such that for every $i=1,\dots,p$,
  $\big\|M^{\vartheta_a}(g_i-\eta_i) \big\|_{\psi_a\mid \mathbf X,\mathbf Y}\leq K_g$, where $\vartheta_a=\min\{1/2,1/a\}$ and the conditional Orlicz norm $\|\cdot\big\|_{\psi_a\mid \mathbf X,\mathbf Y}$ is defined as
  \(\|\cdot \|_{\psi_a\mid \mathbf X,\mathbf Y}
=\inf\left\{C>0:\; 
\mathbb E\left[\exp\!\left(\frac{|\cdot|^a}{C^a}\right)\,\middle|\, \mathbf X,\mathbf Y\right]
\le 2
\right\}.\)
\end{assumption}

Assumption \ref{ass: 1} bounds the average frequency of two types of ranking errors: $\delta_{np}$ controls how often the true mean order is inverted ($\mu_i > \mu_j$), while $b_{np}$ controls how often the aggregated statistic $g$ reverses the correct order ($\eta_j < \eta_i$ when $\mu_j > \mu_i$). We require that $T$ provides a reasonable measure of variable importance, and that the aggregation $g$ does not significantly degrade the ranking quality relative to $T$. As a comparison, MDS \citep{dai_false_2022} requires the inclusion rate $I_j$ (i.e., aggregated statistics) to satisfy the ranking consistency condition (see their Proposition 2.3), which is a stronger condition $(\sup_{i\in N,j\in S}\mathbb{P}(I_i<I_j)\to 0)$. Assumption \ref{ass: 2} requires that, after the rescaling $M^{\vartheta_a}$, $g_i-\eta_i$ has a bounded conditional Orlicz $\psi_a$ norm for every $i$. In other words, each aggregated score has a controlled tail given $(X,Y)$, so we obtain concentration for each variable. The $\psi_a$ family covers sub-Gaussian ($a=2$), sub-exponential ($a=1$), and more generally sub-Weibull ($a>0$) tails; the choice $\vartheta_a=\min\{1/2,1/a\}$ matches the usual $\sqrt{M}$ rate for averages and a slower rate for heavier tails. We justify this assumption holds in several common cases (see Section S1 of the Supplementary Material for details). 


Before stating the theorem, we assume no ties in the aggregation statistics for simplicity; if any occur, we can inject a random tie-breaking ranking scheme, as is common in the literature (see, e.g., \cite{cai2024asymptotic,kanrar2025model}). Let $\gamma = \min_{i\in N,j\in S}|\eta_i-\eta_j|$ denote the minimum separation between any aggregation statistic in $N$ and any in $S$.

\begin{thm}
\label{thm:finitesample}
Assume Assumptions \ref{ass: 1}-\ref{ass: 2} hold. We assume the power of the base FDR controlling procedure is bounded below by some $\kappa>0$. For any target FDR level $q\in(0,1)$ and any finite $(n,p,M)$, there exist constants $c_a>0$ such that the FDR Stabilizer selection $\hat S^{\mathrm{stab}}_q$ satisfies
\begin{enumerate}
    \item $FDR^{\mathrm{stab}} \leq  FDR^{\mathrm{base}}+\frac{1}{\kappa}\Bigg(2p(\delta_{np}+b_{np}) + \frac{1}{s}+4p \exp\Big\{-c_a M^{a\vartheta_a}\Big(\frac{\gamma}{K_g}\Big)^{a}\Big\}\Bigg),$
    \item $Power^{\mathrm{stab}}\geq Power^{\mathrm{base}}-\Bigg(2p(\delta_{np}+b_{np}) + \frac{1}{s}+4p \exp\Big\{-c_a M^{a\vartheta_a}\Big(\frac{\gamma }{K_g}\Big)^{a}\Big\}\Bigg),$
\end{enumerate}
where $FDR^{\mathrm{stab}}$ and $Power^{\mathrm{stab}}$ denote the FDR and the power of the FDR Stabilizer procedure, respectively. $FDR^{\mathrm{base}}$ and $Power^{\mathrm{base}}$ denote the FDR and the power of the base procedure, respectively.

\end{thm}

The three error terms in Theorem \ref{thm:finitesample} have distinct origins: the term $2p(\delta_{np}+b_{np})$ comes from Assumption \ref{ass: 1}. $\delta_{np}$ reflects the intrinsic difficulty of the base statistics, measuring the probability of misordering signal-null pairs. $b_{np}$ counts discordant pairs in which the aggregated scores $\eta_i$ reverse the ordering of $\mu_i$, which measures misordering introduced by the aggregation function. The term $1/s$ arises from a rounding correction accounting for the selection set size $\bar s=\lceil \frac{1}{M}\sum_{m=1}^M\hat{s}_m \rceil$. The exponential term $4p\exp\{-c_aM^{a\vartheta_a}(\gamma/K_g)^a\}$ comes from a concentration bound for $g_i-\eta_i$ under Assumption \ref{ass: 2}, showing that with bounded Orlicz $\psi_a$ norms the deviations shrink exponentially fast as $M$ grows. 

\begin{cor}
\label{thm_fdr}
    Assume Assumptions \ref{ass: 1}-\ref{ass: 2} hold. For any FDR control level $q\in (0,1)$, we assume that the selection set $\hat{S}_q$ obtained by any base procedure satisfies $\limsup_{n,p\to \infty} FDR^{\mathrm{base}}\leq q$, where $FDR^{\mathrm{base}}$ denotes the FDR of the base procedure, and the power of $\hat{S}_q$ is bounded below by some $\kappa>0$. We consider the regime that $s\to \infty$ as $n,p \to \infty$. If $p\delta_{np} =o(1)$, $pb_{np}=o(1)$, and $M \gtrsim (\log p)^{1/(a\vartheta_a)}$, then the FDR Stabilizer selection $\hat{S}^{\mathrm{stab}}_q$ satisfies 
    \begin{enumerate}
        \item $\limsup_{n,p\to \infty} FDR^{\mathrm{stab}}\leq q$,
        \item $\liminf_{n,p\to \infty}Power^{\mathrm{stab}}\geq \liminf_{n,p\to \infty}Power^{\mathrm{base}}$.
    \end{enumerate}
\end{cor}


The verification of the conditions $p\delta_{np}=o(1)$, $pb_{np}=o(1)$, and Assumption \ref{ass: 2} in specific cases is provided in Section S1 of the Supplementary Material. In particular, we show that under mild regularity conditions, both the knockoff and data splitting procedures satisfy $p\delta_{np}=o(1)$; furthermore, when $g$ is chosen as the mean, the selection probability, or the median, Assumption \ref{ass: 2} holds and $pb_{np}=o(1)$ is satisfied.

Next, we focus on examining the stability of the FDR Stabilizer method from the theoretical perspective. The following theorem sheds light on the role of $M$ in the aggregation algorithm. That is, as $M$ approaches infinity, $\hat{S}_q^{\mathrm{stab}}$ obtained by FDR Stabilizer will no longer be randomized. This implies that by increasing the number of iterations in our procedure, we can attain a more stable and reliable selection set. Note that $\mathbb{E}_*$ ($\mathbb{P}_*$) in the theorem denotes the conditional expectation (probability) given data $(\mathbf{X},\mathbf{Y})$, i.e. $\mathbb{E}_*(\cdot):=\mathbb{E}(\cdot \mid \mathbf{X},\mathbf{Y})$ ($\mathbb{P}_*(\cdot):=\mathbb{P}(\cdot \mid \mathbf{X},\mathbf{Y})$). 
\begin{thm}
\label{thm: stability}
Consider an aggregation function $g: \mathbb{R}^M \to \mathbb{R}$ satisfying the bounded difference inequality with parameters ($L_1,\dots,L_M$): for each index $m=1,\dots,M$, 
\[\left| g(t_1,\dots,t_m,\dots,t_M) - g(t_1,\dots,t_m',\dots,t_M) \right| \leq L_m/M, \quad \forall t_1,\dots,t_M,t_m' \in \mathbb{R}.\] 
Let $g_i^{\infty} := \mathbb{E}_*\left[g\left( T_i^{(1)}, T_i^{(2)}, \dots, T_i^{(M)} \right) \right]$, $\Delta := \min \left\{ |g_i^{\infty} - g_{(\lceil \mathbb{E}_*(\hat{s}_m) \rceil)}^{\infty}|/2 : |g_i^{\infty} - g_{(\lceil \mathbb{E}_*(\hat{s}_m) \rceil)}^{\infty}| > 0 \right\}$ and $\delta := \min \left\{ \lceil \mathbb{E}_*(\hat{s}_m) \rceil - \mathbb{E}_*(\hat{s}_m), \mathbb{E}_*(\hat{s}_m) - \lfloor \mathbb{E}_*(\hat{s}_m) \rfloor \right\}$, then we have
\begin{equation*}
\mathbb{P}_*\left( \hat{S}_q^{\mathrm{stab}} = \left\{ i \in [p] : g_i^{\infty} \geq g_{(\lceil \mathbb{E}_*(\hat{s}_m) \rceil)}^{\infty} \right\} \right) \geq 1 - 2p \exp\left(-\frac{2M\Delta^2}{\max_{m\in [M]}L_m^2}\right) - 2\exp\left(-\frac{2M\delta^2}{p^2}\right),
\end{equation*}
where $g_{(1)}^{\infty} \geq \cdots \geq g_{(p)}^{\infty}$ are the order statistics.
\end{thm}
Theorem \ref{thm: stability} concerns the stability of the selection counts across $M$ randomized replicates, conditional on the observed data $(\mathbf{X},\mathbf{Y})$. Once the data are fixed, the $\hat{s}_m$’s are i.i.d. random variables generated purely by the randomization scheme. This explains why the convergence bound depends on $M$ rather than $n$: while the sample size $n$ influences the test statistics, it is not included in the concentration bound conditional on the data.

\begin{remark}
    The set $ \left\{ i \in [p] : g_i^{\infty} \geq g_{(\lceil \mathbb{E}_*(\hat{s}_m) \rceil)}^{\infty} \right\}$ is completely determined by the conditional expectations $g_i^\infty$ and contains no randomness. The error probability in Theorem \ref{thm: stability} has two exponential terms, reflecting two different sources of variability: The first term $2p \exp\left(-\frac{2M\Delta^2}{\max_{m\in [M]}L_m^2}\right)$ arises from the concentration of $g(T_i^{(1)},\dots,T_i^{(M)})$ around $g_i^\infty$ via McDiarmid’s inequality. The bounded-difference constants $L_m$ quantify the sensitivity of $g$ to the $m$-th coordinate; they may depend on $(n,p)$ and even diverge, but a sufficiently large $M$ offsets this growth. The second term $2\exp(-2M\delta^2/p^2)$ controls the deviation of $\bar s$ from $\lceil\mathbb{E}_*(\hat s_m)\rceil$. If $\hat s_m$ can be bounded by $Cs$ then this term can be sharpened to $2\exp(-2M\delta^2/C^2s^2)$, where $C>0$ is a constant. Increasing $M$ improves both aspects simultaneously: it makes the aggregated statistics more representative and the estimated selection size more stable, ensuring that $\hat S_q^{\mathrm{stab}}$ recovers the deterministic target set with high probability.
\end{remark}

\section{Simulations}\label{sec: simulation}

In this section, we conduct a series of numerical simulations to evaluate the empirical performances of FDR control, power, and stability of the FDR Stabilizer method compared with the existing aggregation methods, such as the MBH by \cite{meinshausen_p_2009},
the derandomized knockoffs (derand$\_$kn) by \cite{ren2023derandomized} and 
the MDS by \cite{dai_false_2022,dai_scale_free_2023}. 
In the variable selection scenarios described in this paper, ``BH'' uniformly refers to single split BH \citep{meinshausen_p_2009}. For the above three different base procedures,
their corresponding FDR Stabilizer versions are denoted as ``stab (BH)",  ``stab (kn)" and 
``stab (DS)" in our simulation results, respectively. 

We consider a linear model  $\mathbf{Y} = \mathbf{X} \boldsymbol{\beta} + \boldsymbol{\epsilon}$, where 
the true coefficient vector $\boldsymbol{\beta}$ is sparse. The true signal set 
$S \subset \{1,\dots,p\}$ of size $s$ is randomly selected, where $p$ is the dimension of the covariates and we set $s=50$. For $i \in S$, $\beta_i$ is randomly drawn from $N(0,\delta\sqrt{\log p/n})$, where $n$ is the sample size and $\delta$ is the signal strength parameter. For $i\in \{1,\dots,p\}\backslash S$, we set $\beta_i=0$. 
The error term is independently generated from the standard normal distribution.
The covariates $\mathbf{X}$ are generated by multivariate normal distributions 
$N(0,\Sigma)$, where we consider two different covariance matrices. (1) $\Sigma$ is a blockwise diagonal Toeplitz covariance matrix same as the simulation setting in \cite{dai_false_2022}, whose specific form is present in the Supplementary Material; (2) $\Sigma$ is a compound symmetry covariance matrix with $\Sigma_{ij}=\rho^{\mathbbm{1}(i\ne j)}$. In both cases, $\rho$ measures the strength of correlations among covariates. In our simulations, we consider various cases of $(n,p)$, $\{n=500, p=500\},\{n=800, p=1000\},\{n=800, p=2000\},\{n=2000, p=800\},\{n=3000, p=500\}$, to assess finite sample performances of different methods. In Section S3.4 of the Supplementary Material, we evaluate the method with simulations based on genetic data.


\subsection{FDR control and power analysis}
\begin{figure}[ht!]
    \centering
    \subfigure[$n=500$, $p=500$]{
    \begin{minipage}[b]{.47\linewidth}
        \centering
        \includegraphics[scale=0.43]{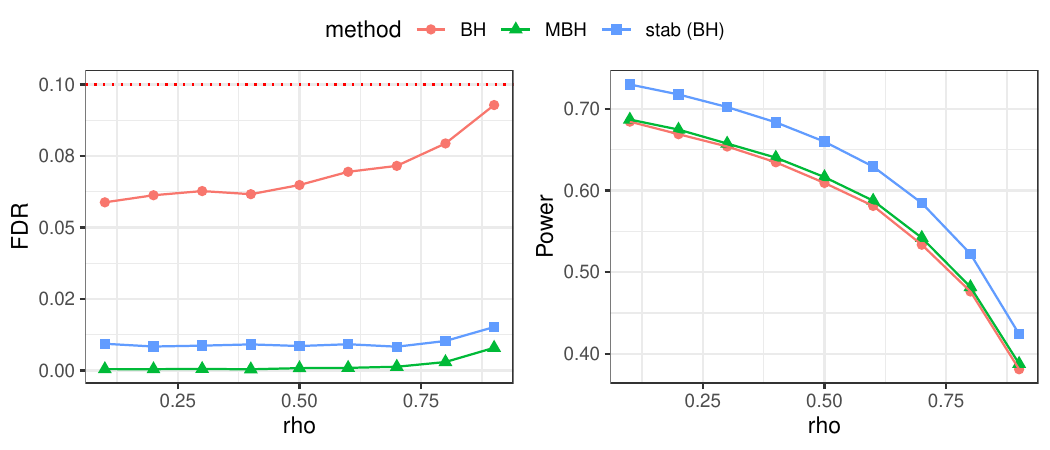}\\
        \includegraphics[scale=0.43]{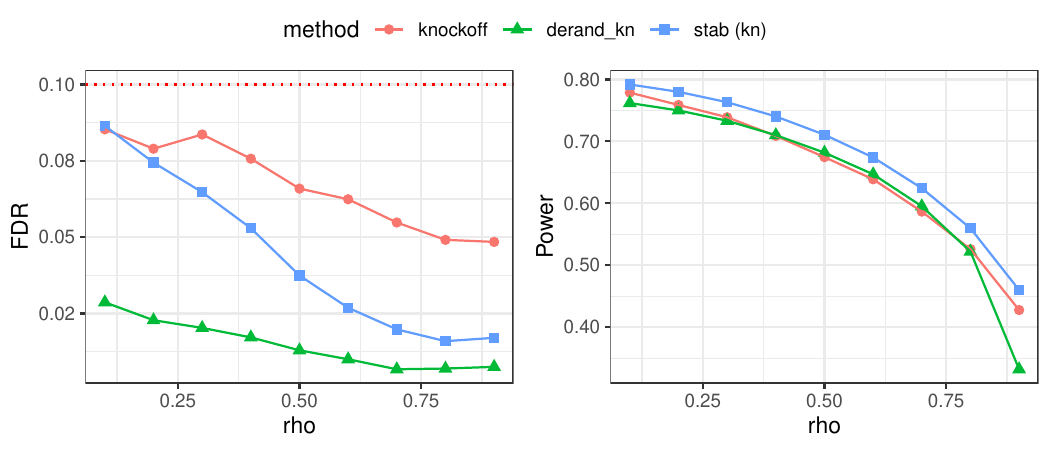}\\
        \includegraphics[scale=0.43]{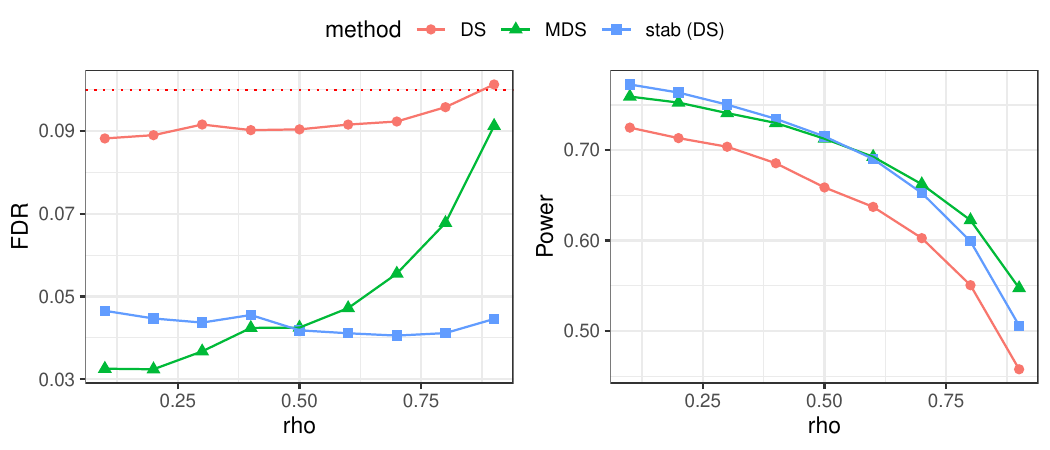}\\
    \end{minipage}
    }
    \subfigure[$n=3000$, $p=500$]{
    \begin{minipage}[b]{.47\linewidth}
        \centering
        \includegraphics[scale=0.43]{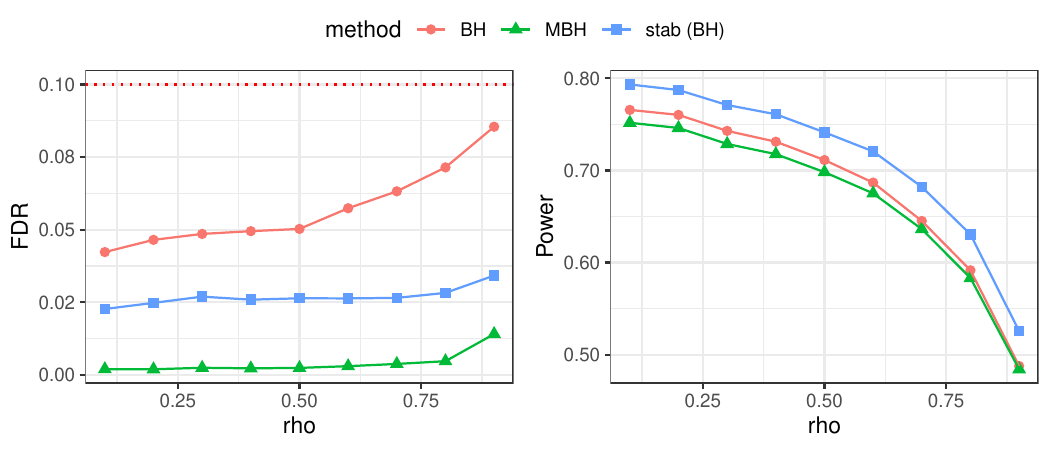}\\
        \includegraphics[scale=0.43]{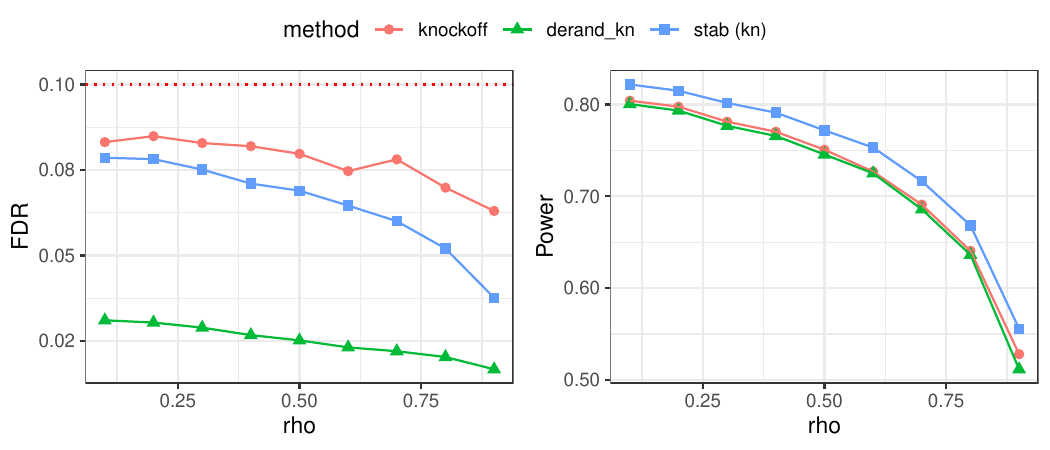}\\
        \includegraphics[scale=0.43]{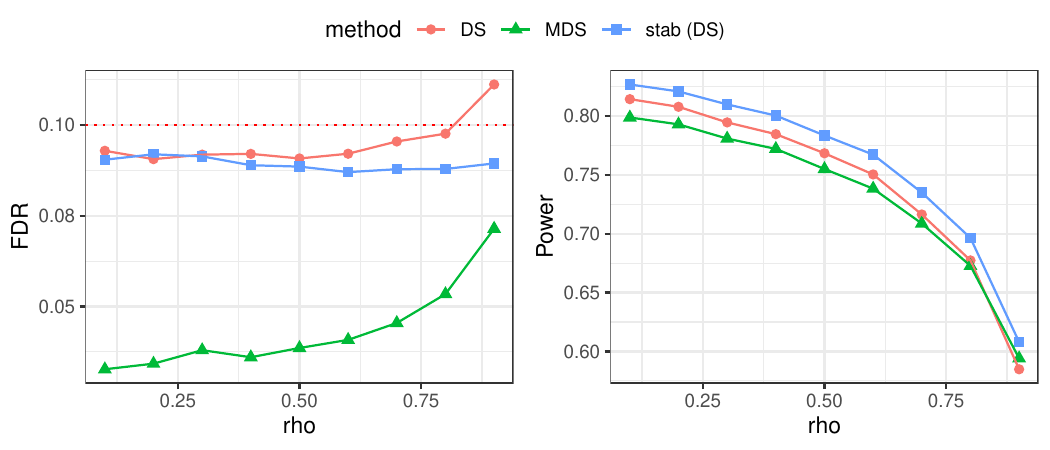}
    \end{minipage}
    }
    \caption{Empirical FDR and power performances of different methods when $\delta=5$, $\Sigma$ is the blockwise diagonal Toeplitz covariance matrix where $\rho$ varies from 0.1 to 0.9. The  specified FDR control level is $q = 0.1$.}
    \label{fig1}
\end{figure}

We first assess the empirical performances of FDR and power. For all aggregation methods, we set the number of independent runs as $M=50$. Figure \ref{fig1} reports the average empirical FDR and power over 1000 independent simulations for the cases $(n,p)=(500,500)$ and $(n,p)=(3000,500)$. In both cases, the signal strength is fixed at $\delta=5$, and $\Sigma$ is the blockwise diagonal Toeplitz covariance matrix where $\rho$ varies from 0.1 to 0.9. The simulation results of other settings are presented in Section S3.1 of the Supplementary Material. 

According to Figure \ref{fig1} (a), we can observe that when the base method is the BH procedure for high-dimensional variable selection in \cite{meinshausen_p_2009}, the FDR Stabilizer method can control the FDR under the given level $q=0.1$ and has better power than the MBH method. 
The middle row of Figure \ref{fig1} shows that when the base method is the Model-X knockoff, the derandomized knockoffs approach appears conservative, resulting in lower power especially when the correlations among covariates are strong. This is because their e-BH procedure depends on the magnitude of the average e-value $\bm{e}^{\mathrm{avg}}$, which is susceptible to the correlations among covariates. To further show this phenomenon, we reduce the signal strength parameter $\delta$ to 2, which is shown in the top panel of Figure \ref{fig: g}. In this case, derandomized knockoff behaved very conservatively, with almost no power. In contrast, the FDR Stabilizer method can enhance the power even for weak signals and strong correlations. 
In the bottom row of Figure \ref{fig1}, we can see that when the base method is Data Splitting in \cite{dai_false_2022,dai_scale_free_2023}, our FDR Stabilizer method and MDS perform relatively close to each other. However, when the correlations among covariates are strong and the signal strength is low, MDS shows a tendency to inflate the FDR. To further show this trend, we reduce the signal strength parameter $\delta$ to 2, which is shown in the top panel of Figure \ref{fig: g}. In this case, MDS can not control FDR. In Figure \ref{fig1} (b), sample size is increased to 3000. In this case, the other three aggregation methods (MBH, MDS, and derandomized knockoff) fail to provide any substantial power improvement over the base procedure, whereas our method achieves a clear and significant power gain.

\subsection{Stability evaluation}

\begin{figure}[ht!]
  \centering
  \includegraphics[scale=0.85]{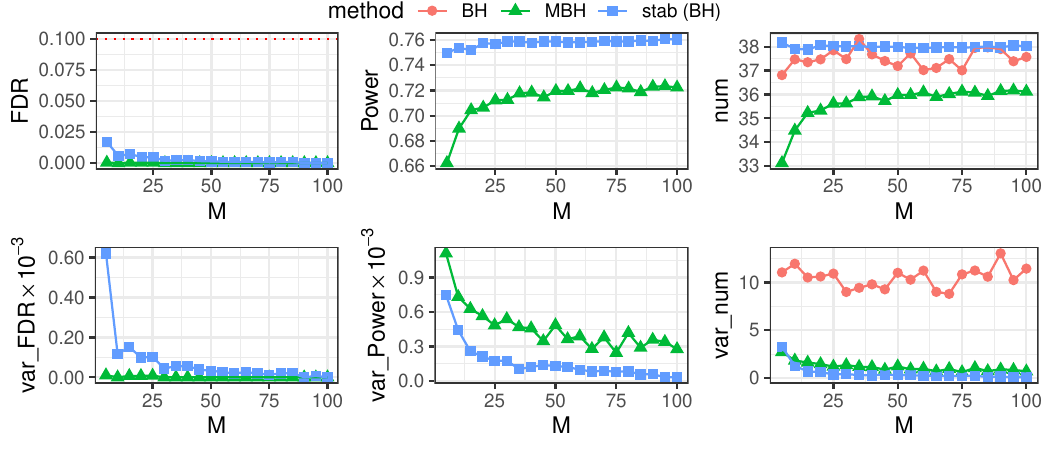}
  \includegraphics[scale=0.85]{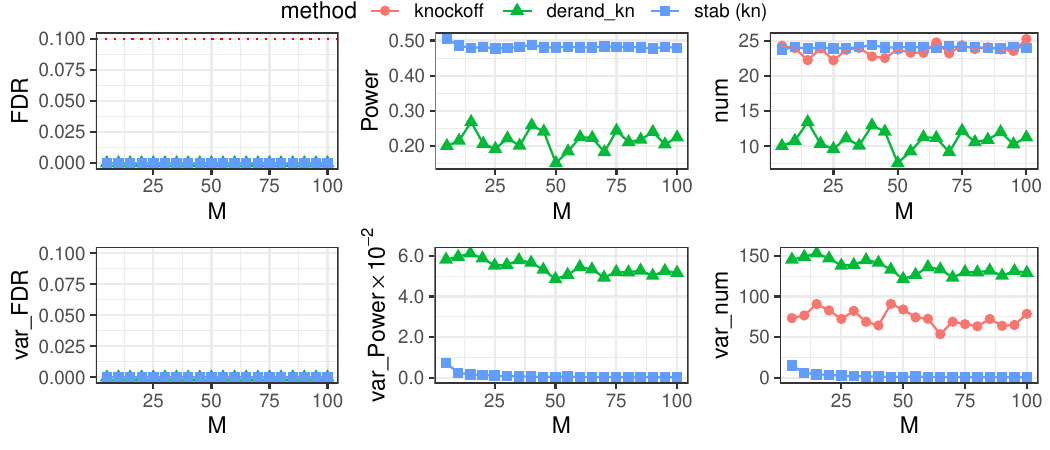}
  \includegraphics[scale=0.85]{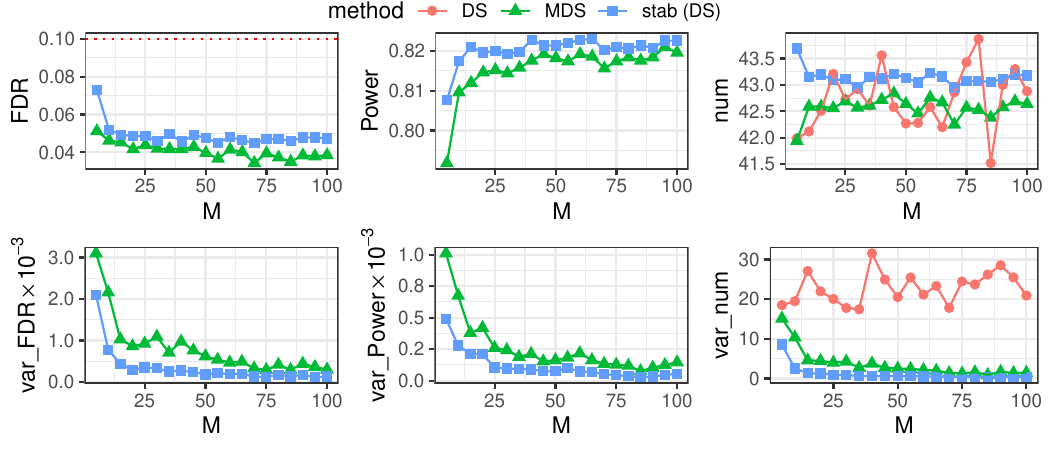}
  \caption{Means (odd rows) and variances (even rows) of FDR, power and number of selected variables for different methods. $n=500, p=500, \delta=7$, $\Sigma$ is the compound symmetry covariance matrix with $\rho=0.4$. The specified FDR control level is $q = 0.1$.}
  \label{fig_stab}
\end{figure}

In this subsection, we compare the stability performance of our method with other aggregation methods. To evaluate stability, we follow a fixed-dataset design: for each simulation scenario, we generate one dataset and keep it fixed throughout the experiment. By this way, randomness originates solely from the FDR control algorithm, not from the dataset. We vary the number of independent runs of the base variable selection procedure, denoted as $M$, from $5$ to $100$ in increments of $5$. For each value of $M$, we repeat the experiment $100$ times to measure the means and the variances of the empirical FDR, empirical power and the number of selected variables, which are reported in Figure \ref{fig_stab} for three aforementioned base procedures when $n=500, p=500, \delta=7$, $\Sigma$ is the compound symmetry covariance matrix with $\rho=0.4$. More simulation results are also included in the Supplementary Material. Our objective is to observe whether there is a tendency for the variances to converge to zero as $M$ increases, the speed at which the variances converge, and whether the change in the means is smooth as $M$ increases.
It is important to note that if the variances of empirical FDR and power converge faster, it implies that the method can achieve stable results with less computational burden.

Figure \ref{fig_stab} demonstrates the excellent stability performances of our FDR Stabilizer method. Our method consistently controls FDR under the nominal level and meanwhile achieves higher average powers with lower variances compared to other aggregation methods. As the number of repetitions $M$ increases, the mean values of FDR, power, and the number of selected variables of the FDR Stabilizer method change smoothly with less fluctuation, significantly improving the stability of the base method. In addition, the variances of FDR, power, and the number of selected variables of the FDR Stabilizer method are consistently lower than other methods and also converge to zero faster, suggesting that we do not need a large $M$ to achieve the stability and save computational time. 
Specifically, rows 3-4 of Figure \ref{fig_stab} illustrate the challenges faced by the derandomized knockoffs approach when dealing with strong correlations within the design matrix. In this case where the correlation parameter $\rho$ is 0.4, the stability performance of derandomized knockoffs is not satisfactory. In contrast, our method effectively stabilizes the base method in the presence of strong correlations. Overall, the simulation results validate the stability property in Theorem \ref{thm: stability} and show that the FDR Stabilizer method is a general stability approach for FDR control in variable selection problems. 

\subsection{Simulation based on genetic data}

The genome-wide association study (GWAS) is a popular approach for genetic research to discover genetic markers associated with a particular phenotype or risk of disease. These genetic markers are often represented as single nucleotide polymorphisms (SNPs), which serve as covariates for variable selection. The dimension of these SNPs is generally larger than the sample size, making the selection of important SNPs challenging.
To mimic a real GWAS study to evaluate the FDR, power, and stability performance of various methods, we use real data as the design matrix $\mathbf{X}$ and randomly generate the response variable $\mathbf{Y}$ based on a linear model as \cite{xing_controlling_2021}. The real dataset contains 292 tomato varieties and 9381 SNPs. Following the steps in \cite{xing_controlling_2021}, we randomly select 1000 SNPs as $\mathbf{X}$, randomly generate $s\in \{60,80\}$ non-zero regression coefficients from $N(0,2500/n)$, and generate the response $\mathbf{Y}$ by $\mathbf{Y} = \mathbf{X} \boldsymbol{\beta} + \boldsymbol{\epsilon}$. The error term is independently generated from the standard normal distribution. We calculate the empirical FDR and power based on $100$ replications with the nominal FDR level $q = 0.1$. Note that the dataset $X$ is fixed; randomness arises from the generation of $\mathbf{Y}$ and from algorithmic randomization.

\begin{figure}[ht!]
  \centering
  \includegraphics[scale=0.46]{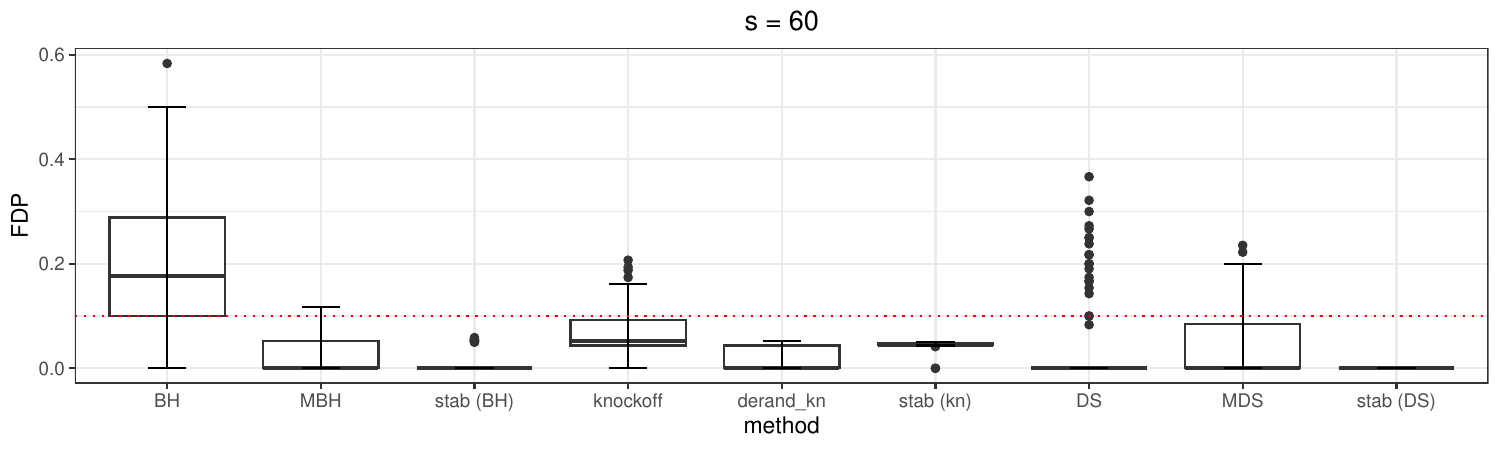}
  \includegraphics[scale=0.46]{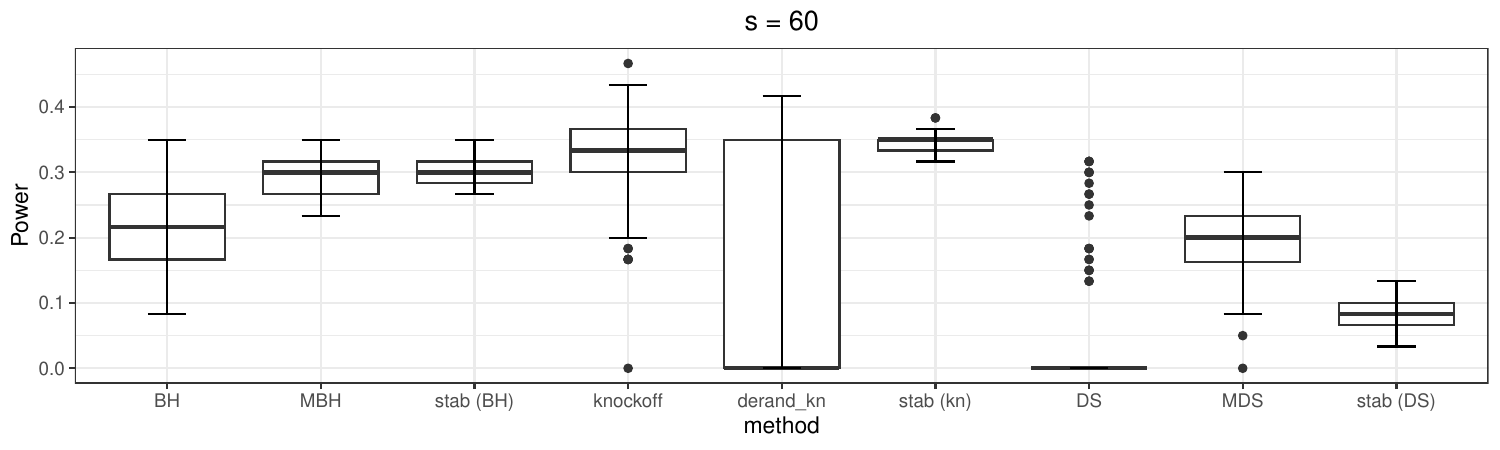}

  \caption{Boxplots of empirical FDR (FDP) and power for the GWAS-based design matrix, over 100 independent trials, with a specified FDR control level of $q = 0.1$.}
  \label{fig_real_data_snp}
\end{figure}

Figure \ref{fig_real_data_snp} provides compelling evidence that our method can consistently control the FDR while achieving higher power than the base and other aggregation methods. Meanwhile, the boxplots show that our FDR Stabilizer method achieves lower variances of empirical FDR and powers, which indicates that our method can significantly improve the stability of the base variable selection methods and is also more stable than other existing aggregation methods. It is widely recognized that genetic data often exhibit strong correlations, and this example emphasizes the findings observed in previous simulations. Specifically, for MDS, strong correlations and weak signal may lead to inflated FDR. The derandomized knockoffs method shows an unstable tendency in the case of $s = 60$, and when $s = 80$, the power is even reduced to 0. In contrast, our method not only enjoys excellent finite sample  performances in terms of FDR and power but also proves to be highly competitive in terms of stability. It is important to note that SNPs usually have only three possible values, making the design matrix in GWAS studies  challenging for regression problems. This results in low powers and unstable selections for the base method. These results further validate the effectiveness and reliability of our approach applied to genetic data.

\section{Applications to real data}\label{sec: real data}

\subsection{HIV drug resistance}

\begin{figure}[ht!]
  \centering
  \includegraphics[scale=0.46]{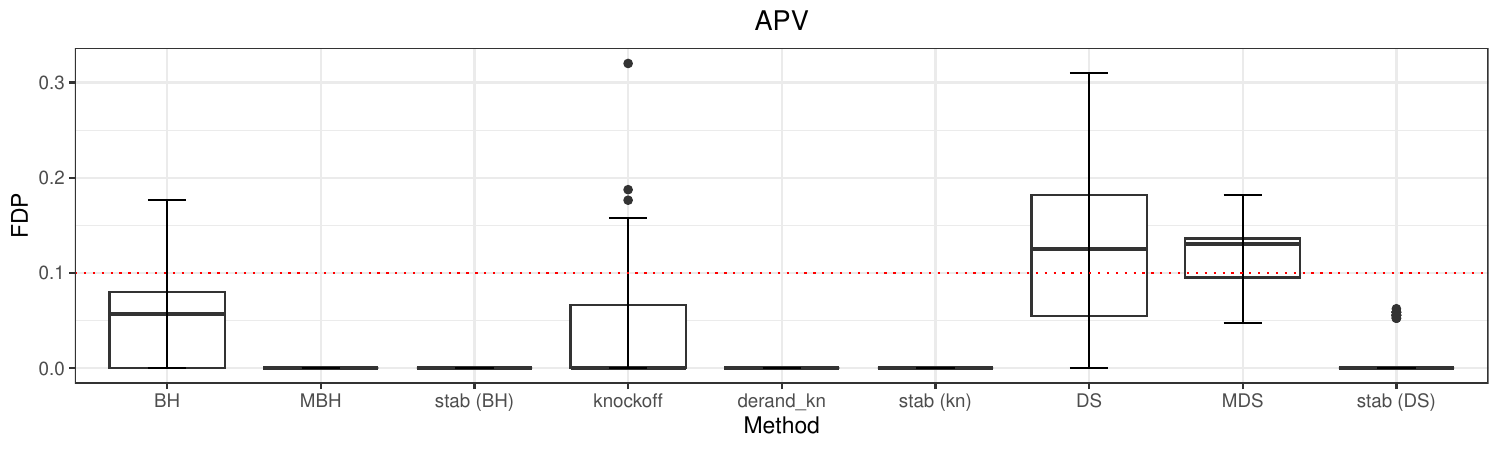}
  \includegraphics[scale=0.46]{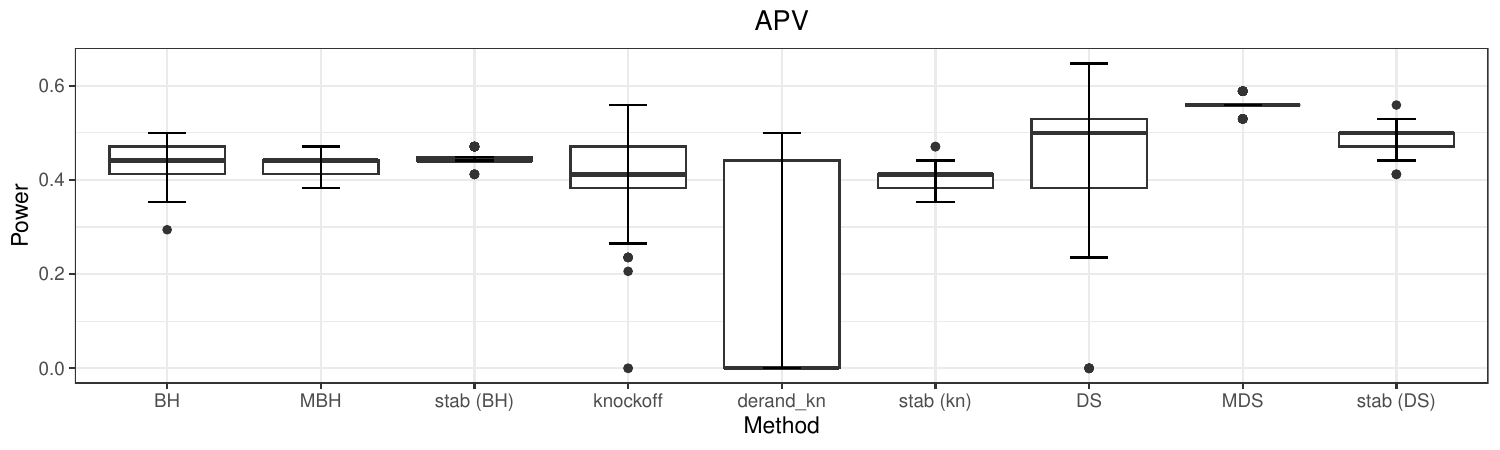}
  \caption{Boxplots of empirical FDR (FDP) and power for various methods over 100 independent trials for the HIV Drug Resistance data.}
  \label{fig_real_data}
\end{figure}

We next compare the empirical performance of three base methods, BH, Model-X knockoffs, Data Splitting and their stabilized versions, using the well-known data \citep{HIV} for detecting mutations associated with drug resistance in human immunodeficiency virus type 1 (HIV-1). This dataset contains resistance measurements for seven protease inhibitors (PIs) drugs, six nucleoside reverse transcriptase inhibitors (NRTIs) drugs, and three non-nucleoside reverse transcriptase inhibitors (NNRTIs) drugs. For the sake of brevity, we only focus on PIs. 
To deal with missing data and preprocess the dataset, we mainly follow the steps in \cite{barber_controlling_2015}. For sample $i$ and mutation $j$, $Y_i$ denotes the logarithm of the increase in resistance to the drug, and the design matrix $\mathbf{X} = (X_{ij})\in \{0,1\}^{n\times p}$ contains binary variables. If mutation $j$ is present in sample $i$, then $X_{ij}=1$. We set the target FDR level $q=0.1$, and apply nine methods based on BH, knockoff, and DS to detect the mutations in the HIV-1 associated with resistance to each drug. For derandomized knockoffs, we set $q=0.1$ and $q_{\mathrm{kn}}=0.05$. For each aggregation method, the number of repeated runs is $M = 50$. The dataset itself is fixed, but the randomness comes from the FDR control procedure. We therefore repeat the complete analysis 100 times on this single dataset to evaluate FDR and power. The results of this analysis are presented in Figure \ref{fig_real_data}.

To evaluate the performance of FDR and power, we treat the existing treatment-selected mutation (TSM) panels as the ground truth. The boxplots provided in Figure \ref{fig_real_data} offer a visual representation of the empirical FDR and power for each method. Here we only show two types of PI drugs, APV and LPV, the rest can be found in Section 4.1 of the Supplementary Material. The concentration trend from these boxplots shows that our proposed method exhibits excellent stability. In addition, our method controls the FDR well and has a relatively high power. It is important to note that derandomized knockoff tends to be overly conservative and unstable in this highly correlated scenario, which aligns with the findings from the previous simulations. This conservatism translates into low power or even no power in certain instances, limiting its effectiveness in detecting significant mutations.
Furthermore, although MDS appears to have higher power compared to other methods, it often fails to control the FDR below the target level. This compromises its reliability and suitability to detect drug resistance-associated mutations in HIV-1.
Overall, our method not only achieves stability but also maintains a balance between FDR control and power, making it a highly competitive approach for this research domain.

\subsection{Test predictability of proteins in CITE-seq data}

CITE-seq is a recent multimodal single-cell phenotyping technology that contains measurements of single-cell gene expression and surface proteins. While the gene expression data are high-dimensional, noisy, and sparse, the surface proteins are typically low-dimensional, highly informative, and expensive to measure. As a result, predicting the surface proteins based on gene expression data provides better understanding of the RNA translation process, and enables researchers to obtain the estimated protein levels when only the RNA sequence is measured \citep{protein_2020}. \cite{cai_model-free_2022} proposed a model-free prediction test that incorporates the strategy of sample splitting. The authors applied the Cauchy combination test \citep{liu2020cauchy} to aggregate the dependent $p$-values obtained from multiple random data splits. Here we modify their aggregation method to fit in our FDR Stabilizer
framework and compare the stability of the results. Specifically, we treat the gene expression data as $\mathbf{X}$ and treat one specific protein as $\mathbf{Y}$. The task is to test
\[H_{0j}:\mathbb{E}(Y_j)=\mathbb{E}(Y_j|\mathbf{X})\quad\mathrm{vs.}\quad H_{1j}:\mathbb{E}(Y_j)\neq\mathbb{E}(Y_j|\mathbf{X}), j\in [p].\]
We apply the same quality control procedures as in \cite{cai_model-free_2022} to obtain the marker genes and use the same machine learning models and two-sample tests to obtain the $p$-values, which are based on a single random data split. With multiple splits, the $p$ values are then aggregated using the Cauchy combination test. 
To control the FDR, \cite{cai_model-free_2022} utilized the BY method \citep{benjamini2001control} on the aggregated $p$-values. To incorporate their approach into our framework, we consider the base FDR control procedure as running the BY procedure under the FDR control level of $q=0.1$ on the $p$-values obtained from a single split. 
The algorithm details are available in Section S4.2 of the Supplementary Material.

Our analysis reveals that both methods identify similar proteins with predictive potential, as detailed in Section S4.2 of the Supplementary Material. To assess the stability, Table \ref{tab1} summarizes the variances of the rejections of the null hypotheses for the eight cell types from 20 independent runs, where each run consists of a combination of 50 random splits. ``CCT'' and ``stab'' in the table header denote the aggregation method using Cauchy combination and our aggregation method, respectively. The suffixes ``all'' and ``marker'' indicate the use of top 5000 highly variable genes as X and the use of marker genes as X, respectively. It shows that our method exhibits lower variances and demonstrates higher stability compared to the method in \cite{cai_model-free_2022}.

\begin{table}[ht]
\centering
\renewcommand{\arraystretch}{0.85}
\begin{tabular}{ccccc}
\hline
Cell type & CCT all & CCT marker & stab all & stab marker \\ 
\hline
Mono   & 0.516 & 0.766 & 0.263 & 0.253 \\
B      & 0.555 & 0.345 & 0.000 & 0.000 \\
CD4 T  & 0.345 & 0.576 & 0.221 & 0.221 \\
CD8 T  & 0.937 & 0.368 & 0.155 & 0.239 \\
NK     & 0.682 & 0.766 & 0.263 & 0.239 \\
DC     & 1.503 & 0.976 & 0.263 & 0.168 \\
Other  & 0.747 & 0.568 & 0.261 & 0.000 \\
Other T& 1.579 & 0.892 & 0.197 & 0.000 \\
\hline
\end{tabular}
\caption{The variances of the rejections of the null hypotheses 
for the eight cell types based on $20$ independent replications, where each replication consists of 50 random splits.}
\label{tab1}
\end{table}

\section{Discussion}

In this paper, we propose a general stability framework to aggregate the results of multiple runs of FDR control methods. The proposed FDR Stabilizer approach addresses the inherently stochastic nature of FDR control methods, providing a new stable way for FDR control and can avoid power loss. Our method aggregates statistics generated from multiple runs of the base algorithm to construct stabilized e-values, which are then processed using the e-BH procedure. Notably, our approach is the first to explicitly connect the stabilization of FDR control with rank aggregation, thereby opening up a new perspective that bridges multiple testing with the broader literature on ranking and consensus methods. In practical applications, our proposed method exhibits exceptional stability, enhancing the power of the base method while simultaneously controlling FDR compared to other commonly employed aggregation methods. 

Furthermore, when an algorithm lacks stochasticity—implying that its modeling output remains constant for a given dataset—our approach can be employed to bolster algorithmic stability through data perturbation techniques, such as subsampling and bootstrapping. For high-dimensional variable selection challenges, \cite{meinshausen_stability_2010} proposed an innovative stability selection method that integrates subsampling with selection algorithms. To illustrate how FDR Stabilizer performs under data perturbation, we carry out a perturbated multiple test experiment in Section S3.3 of the Supplementary Material, where we show that the same theoretical guarantees remain valid and that power is preserved. This extension underscores the broad applicability of our proposed methodology. 

{

\bibliographystyle{apalike}
\bibliography{Bibliography-MM-MC}
}

\end{document}